\documentclass[aps,prl,twocolumn,amsmath,amssymb,longbibliography]{revtex4-1}

\usepackage[utf8]{inputenc}
\usepackage{graphicx}
\usepackage{booktabs}
\usepackage{notes2bib}

\usepackage[colorlinks=true,bookmarks=false,citecolor=blue,urlcolor=blue,pdfstartview=FitH]{hyperref}

\begin{document}

\title{Topological waves guided by a glide-reflection symmetric crystal interface}
\author{Julio Andrés Iglesias Mart\'inez, Nicolas Laforge, Muamer Kadic, and Vincent Laude}
\affiliation{Institut FEMTO-ST, CNRS UMR 6174, Université Bourgogne Franche-Comté, Besançon, France}

\begin{abstract}
A domain wall separating two different topological phases of the same crystal can support the propagation of backscattering-immune guided waves.
In valley-Hall and quantum-Hall crystal waveguides, this property stems from symmetry protection and results from a topological transition at a Dirac point.
Since an initially closed band gap has to open, the guidance bandwidth remains limited compared to that of wide band gap crystals.
When a glide-symmetric dislocation is introduced in a 2D crystal, we show that a pair of wide-bandwidth, single-mode, and symmetry-protected guided waves appear in the bulk band gap.
The 2D Zak phase changes by $\pi$ on either side of the interface, providing a topological invariant protected by glide-reflection symmetry at the X point of the Brillouin zone.
A demonstration experiment is performed with acoustic waves in water, at ultrasonic frequencies, and shows the continuous tuning of transmission as a function of the glide parameter.
The concept further extends to other types of waves, including the case of elastic waves in solids, but also of optical and electromagnetic waves.
\end{abstract}

\maketitle

\section{Introduction}

Topological phononics promises unprecedented wave properties inspired by the concepts of topological insulators~\cite{hasanRMP2010,fleuryS2014,yangPRL2015,zhangCP2018,maNRP2019,gaoNN2021}.
One promising direction is the achievement of uni-directional and backscattering-free guided wave propagation along a boundary of a crystal or a domain wall between two crystal phases.
Passive topological waveguides, for instance of the valley-Hall~\cite{luNP2017,yanNM2018} and the quantum-Hall type~\cite{wuPRL2015,heNP2016,miniaciPRX2018}, lead to symmetry-protected, single-mode guided waves along a domain wall separating two phases of the same crystal with different topological invariants.
The topological properties of the waveguide are inherited from those of the two-dimensional (2D) bulk crystal according to the bulk-boundary correspondence principle~\cite{mongPRB2011}.
The one-dimensional (1D) domain wall hetero-structure is formed without breaking the periodicity of the 2D lattice, by tuning a continuous geometrical parameter that controls a topological transition~\cite{luPRB2014,chen_topological_2019}.
For instance, in valley-Hall crystals, a triangular inclusion is rotated continuously to reduce the symmetry of a 2D crystal possessing a band structure with a Dirac point at the K point of the first Brillouin zone, causing a gap to open there~\cite{luPRL2016,luNP2017,zhuPRB2018}.
In quantum-Hall crystals with C6v symmetry, a double Dirac at the $\Gamma$ point undergoes a topological transition under the continuous tuning of the internal structure of the unit cell of the crystal~\cite{wuPRL2015,heNP2016,palNJP2017}.
In both cases, however, the available bandwidth for the dispersion of the guided wave is limited by the effective opening of the band gap that the control parameter allows~\cite{laudeAPLM2021}.
In contrast, artificial crystals have long been designed to present very wide complete band gaps~\cite{vasseurPRL2001,laudeBOOK2020,iglesiasAPL2021} that the guided bands could in principle cover.
Phononic crystal waveguides formed by coupling a sequence of crystal defects, however, lack topological protection and are generally multimodal, leading to a competition of the guided bands inside the complete band gap that can severely flatten the guided bands~\cite{laudeAPLM2021}.

Can we obtain topological crystal waveguides that make full use of a wide complete band gap crystal?
We propose in this Letter to start from a 2-periodic crystal and to introduce a glide-reflection (GR) symmetric dislocation running all along one of the periodicity axes.
The resulting structure loses one periodicity, along the directional orthogonal to the glide operation, but gains a glide-reflection symmetry that the initial 2-periodic crystal did not possess.
The 2D Zak phase of bulk bands, measured along the interface direction, changes by $\pi$ on either side, providing a topological invariant protected by GR symmetry.
All pairs of bands at the boundary of the first Brillouin zone (X point) are degenerate, leading to the appearance of pairs of left- and right-propagating guided waves in all Bragg band gaps of the crystal.
The pairs of guided Bloch waves are protected by GR symmetry at the X point and their smooth dispersion covers most of the band gap of the bulk crystal.

The Letter is organized as follows.
We first discuss the topology of the band structure of a square-lattice crystal and its transformation under a glide dislocation.
We show how the dispersion of waves guided along the glide dislocation closes the complete band gap exactly for a half-lattice glide dislocation.
Tuning the glide parameter, the spectral transmission can be changed continuously from no transmission at all to full transmission through the phononic band gap.
The glide-reflection symmetric crystal waveguide offers wide bandwidth, single mode operation, and symmetry-protected backscattering immunity.
An experiment performed with ultrasonic acoustic waves around 0.5 MHz and a crystal of steel rods in water demonstrates the operation of the glide-reflection symmetric phononic crystal waveguide.

\begin{figure}[t]
\centering
\includegraphics[width=75mm]{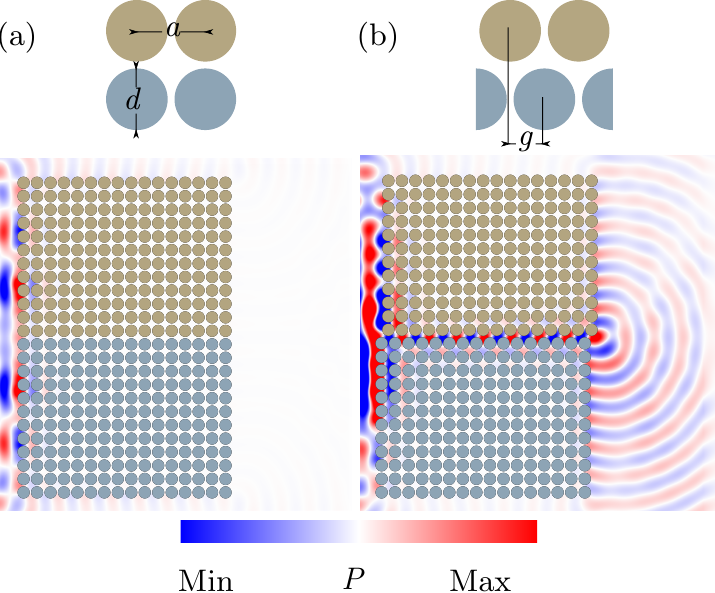}
\caption{Principle of the glide-reflection symmetric topological phononic crystal waveguide.
(a) A 2-periodic square-lattice phononic crystal is composed of steel rods in water (lattice constant $a$, diameter $d=0.9a$).
For every frequency within the complete band gap, transmission of incident acoustic waves is forbidden.
(b) Two pieces of the same square-lattice crystal are now glided along the $x$-axis.
The glide parameter $g$ is periodic with period $a$.
For a half-lattice glide parameter ($g=a/2$), waves are guided along the glide dislocation, for frequencies within the complete band gap.
In numerical simulations, $P$ is the normalized pressure field, frequency is taken at the center of the band gap, and waves are incident from the left.}
\label{fig1}
\end{figure}

\bigskip

For demonstration purposes, we consider in the following a 2-periodic square-lattice phononic crystal of circular inclusions, as shown in Figure~\ref{fig1}.
Numerical simulations are performed in this Letter considering a phononic crystal of steel rods in water, but the results extend naturally to other material systems \bibnote[SM]{See Supplemental Material at \url{link_to_be_defined} for more information on other material systems, for a discussion of modal symmetry, for experimental details, and for exploration of resilience to disorder.}.
In the glide dislocation, one half of the crystal is spatially shifted by an amount $g$, along direction $x$.
The glide operation creates an interface between two crystal phases that remain identical except for the spatial shift.
The initial crystal ($g=0$) possesses a complete band gap within which transmission decreases exponentially with crystal thickness.
Fig.~\ref{fig1}(a) illustrates numerically, for a frequency at the center of the complete band gap, that total reflection of incident waves results.
When $g=a/2$, guided waves appear along the dislocation and transmission is obtained, as Fig.~\ref{fig1}(b) shows.

\begin{figure}[bt]
\centering
\includegraphics[width=80mm]{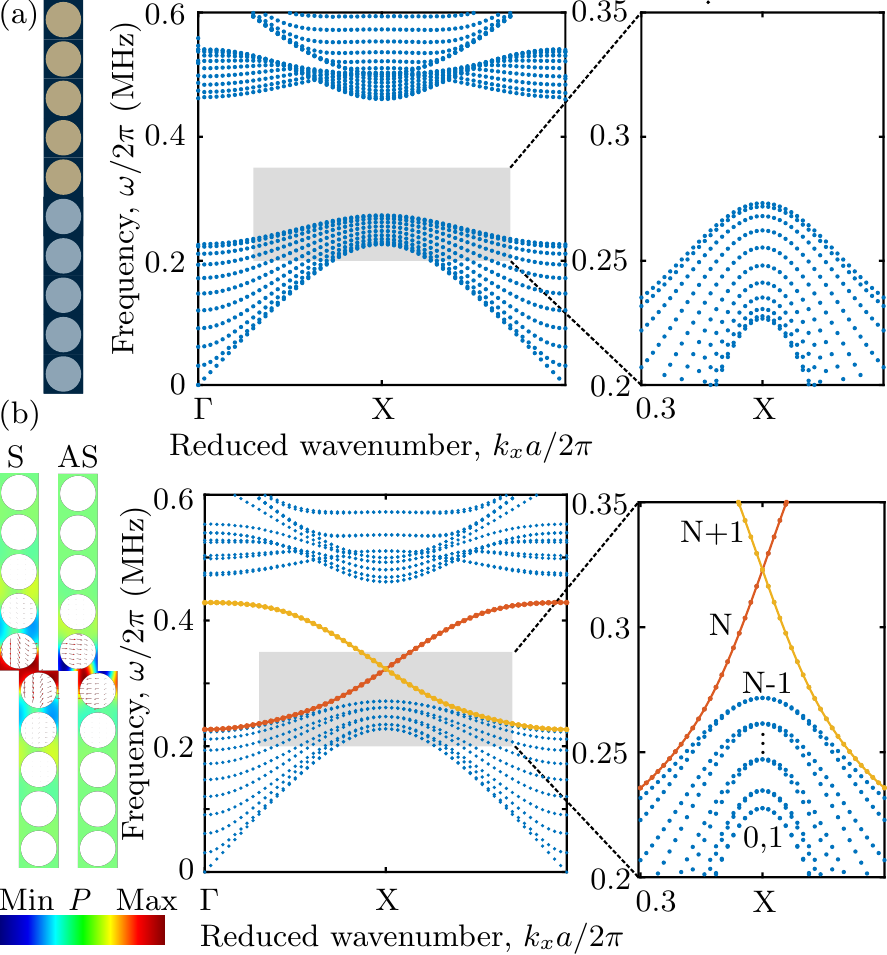}
\caption{Band structure topology of the glide-reflection waveguide, computed for a supercell made of $N=10$ unit cells of the crystal.
(a) For $g=0$, the supercell simply repeats $N$ times vertically the primitive cell of the 2D square-lattice crystal ($d=2$ mm, $a=2.22$ mm).
The band structure of the waveguide is obtained from the projected band structure of the 2D crystal.
(b) For $g=a/2$, the bands group by pairs of symmetric (S) / anti-symmetric (AS) Bloch waves, with respect to the glide reflection symmetry.
They are degenerate at the Brillouin zone edge (the X point), causing a pair of guided waves to appear inside the complete band gap.
The modal distributions of the S (red color band) and AS (yellow color band) guided waves are shown on the left for $k_x a / \pi = 0.8$.
}
\label{fig2}
\end{figure}

Let us analyze the topology of the band structure of the phononic crystal structure and its change with the glide parameter.
Figure \ref{fig2}(a) shows a super-cell and the phononic band structure for the 2-periodic crystal for $g=0$; Fig. \ref{fig2}(b) shows similar information for the glide-reflection symmetric waveguide for $g=a/2$.
The super-cell is the numerical device used to obtain the dispersion relation of crystal waveguides.
Periodic boundary conditions are applied on the vertical boundaries of the supercell while the horizontal boundaries are left free (Neumann boundary condition).
In the figure, we consider $N=10$ unit cells in the vertical direction and $1$ unit cell in the horizontal direction.
The band structures account for acousto-elastic coupling between pressure waves in water and elastic waves in steel \cite{laforgePRA2021}.
The band structure for $g=0$ in Fig. \ref{fig2}(a) shows the complete band gap separated by groups of bands.
Counting the bands, there are exactly $N+1$ bands below the band gap.
Those bands are actually sampled from the original Brillouin zone as $(k_ya/\pi=n/N, k_x)$ with $n=0, \cdots, N$~\cite{laudeBOOK2020}.
Hence, when $n=N$, the Bloch wavevector varies along the YM edge of the first Brillouin zone ($k_ya/\pi=1$).

The guided waves for $g \neq 0$ appearing inside the band gap originate from the $N$-th and $(N+1)$-th bands.
Actually, as Fig. \ref{fig2}(b) shows for glide parameter $g=a/2$, all bands are degenerate by pairs at the X point of the first Brillouin zone.
This essential property is obtained only for a half-lattice glide; it is shown later to signal a topological transition of the band structure occurring at $g=a/2$.
Since the $N$-th and the $(N+1)$-th bands were repelling and thus sitting on opposite sides on the band gap for $g=0$ and they are degenerate at the X point for $g=a/2$, they have to move inside the band gap as $g$ is tuned continuously between those two values.
When $g>a/2$ and is tuned toward $g=a$, the gap closes continuously and symmetrically from the case $g<a/2$.

Why the $N$-th and $(N+1)$-th bands hold a pair a guided Bloch waves can be understood based upon the transformation of the band structure under the continuous change of glide parameter $g$ from $0$ to $a/2$.
The dispersion of the guided wave extends inside the complete band gap, with a real wavevector $k_x$; along the $y$ direction the guided wave is evanescent,  i.e. its amplitude is decreasing exponentially.
Fig. \ref{fig2}(b) illustrates that property for one particular value of $k_x$.
In contrast to the case $g=0$, the Bloch waves of the 2-periodic phononic crystal do not translate directly into Bloch waves of the waveguide structure for $g \ne 0$.
However, they can still be used as a functional basis to express the $1$-periodic guided mode.
Bloch waves of the 2-periodic crystal are all evanescent for frequencies inside the complete band gap.
Hence their wavevector satisfies $\Re(k_y)a/\pi=1$: the real part of the Bloch wavevector is restricted to the top edge of the 2D first Brillouin zone.
Furthermore, the imaginary part of the Bloch wavevector can only be directed along the $k_y$ direction in reciprocal space, since propagation is lossless along the $x$-axis.
$\Im(k_y) \neq 0$ hence provides the necessary exponential decrease away from the glide interface such that the guided wave is confined.
It can be check visually in the modal shapes of Fig. \ref{fig2}(b) that the exponential decrease in amplitude along the $y$-axis is accompanied by an alternation of the sign from one unit cell to the next, in accordance with the condition $\Re(k_y)a/\pi=1$.

Zak phase was originally introduced \cite{zakPRL1989} for 1D crystals as the integral of the Berry connection along the 1D Brillouin zone.
The 2D Zak phase for 2D crystals \cite{liuPRL2017,liuPRB2018} is a natural generalization where the integral of the Berry connection is taken along a 1D contour, chosen as the interface direction in reciprocal space \cite{maNRP2019}.
Namely, Zak phase for band $n$ is
\begin{equation}
\gamma_n = \int_C \rm{d}{\bf R} \cdot {\cal A}_n({\bf R}) ,
\end{equation}
with ${\bf R}=k_x$ at fixed $k_y$ and $C=[-\pi/a;\pi/a]$.
The Berry connection is
\begin{equation}
{\cal A}_n({\bf R}) = i \langle u_n({\bf R}) | \nabla_{\bf{R}} | u_n({\bf R}) \rangle ,
\end{equation}
with $u_n({\bf R})$ a Bloch wave defined over the 2D unit cell and $\langle . \rangle$ denoting the scalar product in real space defined on this unit cell.
Note that the integration contour chosen, $C$, is different from the one used to define Chern numbers, that is the boundary enclosing the first 2D Brillouin zone.
By construction, the bottom crystal B is the glide-reflection (GR) image of crystal A.
The glide operation implies a phase change for every Bloch wave $\phi(k_x)=-gk_x$ (a translation of the origin by $g$).
Since the Berry connection changes as ${\cal A}_n({\bf R}) \rightarrow {\cal A}_n({\bf R}) - \frac{\partial \phi}{\partial k_x}$ under any phase change, we have
\begin{equation}
\gamma_n(B) = \gamma_n(A) + 2\pi g/a .
\end{equation}
Hence there is a $\pi$ change of the 2D Zak phase across the interface for every band, for $g=a/2$ exactly.
Since the Zak phase is $2\pi$-periodic, its value alternates by $\pi$ between both crystal images.

Why degenerescence of Bloch waves occurs specifically for $g=a/2$ and at the X point of the first Brillouin zone results from the combination of the space group symmetry of the waveguide and of its periodicity along the glide dislocation, as the Supplemental Material (SM) details \bibnotemark[SM].
Let us consider here a compact demonstration based on operators of the 1D crystal interface.
For any glide parameter $g$, the composition $G_{a-g} \circ G_g$ is the translation by one lattice constant $T_a$ in direct space.
In reciprocal space, this implies $G_{a-g}(k) G_g(k) = \exp(i k a)$.
For $g=a/2$, we then have $G^2_g(k) = -1$ at the X point of the 1D Brillouin zone ($k a = \pi$).
Hence the eigenvalues of $G_g(\pi/a)$ are $\pm i$.
Its eigenvectors form complex conjugate pairs, since $G_g(\pi/a)u=iu$ implies $G_g(\pi/a)u^*=-iu^*$.
Since the glide operator commutes with the dynamical operator for the wave equation, they share common eigenvectors.
Hence $Du=\omega^2u$ implies $Du^*=\omega^2u^*$ since the wave equation has real coefficients because of its time reversal invariance (TRI).
Summarizing, each complex conjugate eigenvector pair shares a degenerate eigenvalue at the X point.

\bigskip

\begin{figure}[t]
\centering
\includegraphics[width=75mm]{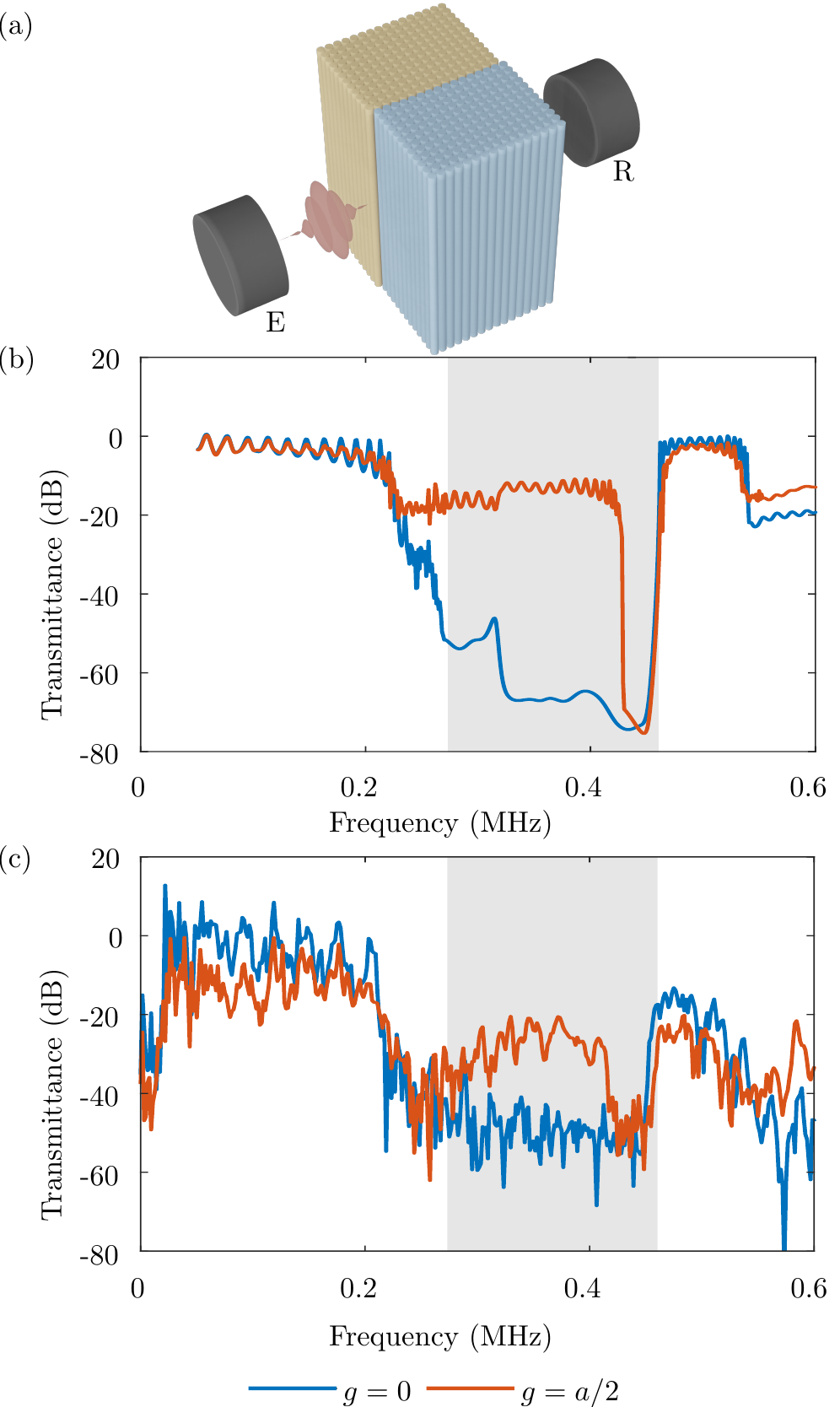}
\caption{Glide-reflection symmetric topological phononic crystal waveguide experiment.
(a) Two pieces of the same square-lattice crystal of steel rods in water (diameter $d=2$ mm; lattice constant $a=2.22$ mm) are glided.
The transmission of acoustic waves guided along the dislocation is probed using an ultrasonic emitter (E) of short pulses that are detected by an ultrasonic receiver (R).
For a half-lattice glide parameter ($g=a/2$), the (b) numerical and (c) experimental acoustic wave transmission covers most of the complete phononic band gap (highlighted with the grey color).
The reference for transmission is the measurement in the absence of the phononic crystal waveguide.}
\label{fig3}
\end{figure}

\begin{figure}[tb]
\centering
\includegraphics[width=75mm]{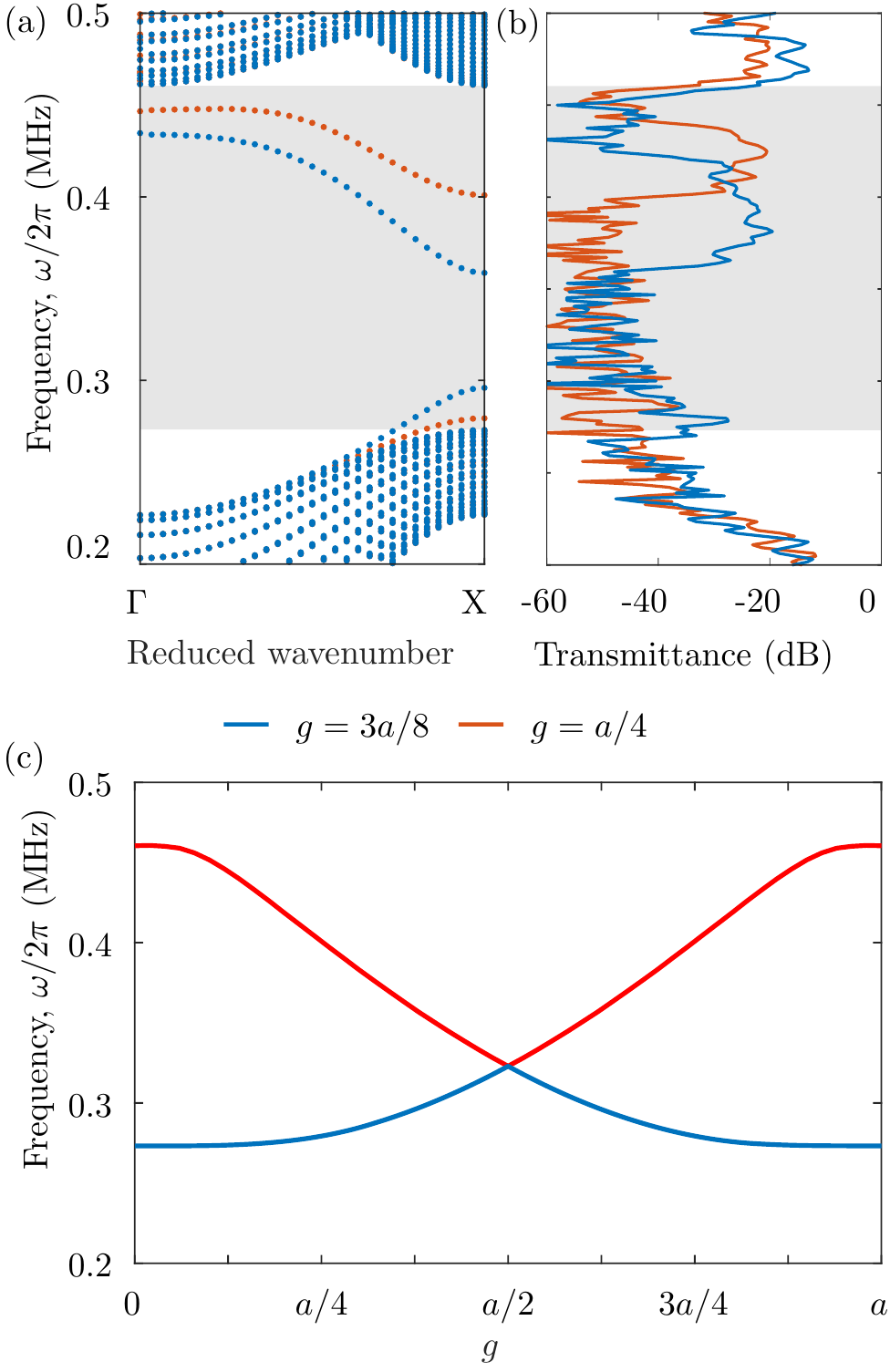}
\caption{Gap opening as a function of the glide parameter.
(a) For $g \neq a/2$, a mini-gap for guided waves opens in the phononic band structure at the X point of the Brillouin zone.
(b) Experiment confirms the opening of the mini-gap, for $g=a/4$ and $g=3a/8$.
(c) The eigenfrequencies of the two guided waves at the X point vary with the glide parameter (blue line: S waves; red line: AS waves).
For exactly $g=a/2$, the waveguide is glide-reflection symmetric and the guided wave gap closes.
This gap opens symmetrically on either side of that value.}
\label{fig4}
\end{figure}

We now turn to the experimental demonstration of the glide-reflection symmetric phononic crystal waveguide.
The phononic crystal of steel rods in water is depicted in Figure~\ref{fig3}(a).
A total of $24 \times 16$ parallel rods are aligned using perforated parallel plates \bibnotemark[SM].
The rod diameter is $d=2$ mm and the lattice constant is $a=2.22$ mm ($d/a=0.9$).
Fig. \ref{fig3}(b) shows the numerical transmission as a function of frequency.
The experimental transmission of Fig. \ref{fig3}(c) is obtained based on the ultrasonic pulse-echo technique described for instance in Ref. \cite{khelifPRB2003b}.
The complete band gap extends from 0.28 MHz to 0.46 MHz whereas the guided mode transmission covers the range from 0.28 MHz to 0.43 MHz, in accordance with theory.

For other glide parameter values, transmission is observed as well inside the complete band gap but within a reduced frequency range.
Actually, for $g \ne a/2$ the left and right propagating guided Bloch waves interfere and form an anti-crossing and thus a mini band gap at the X point, as  Fig.~\ref{fig4}(a) shows.
The experiments reported in Fig.~\ref{fig4}(b) clearly show the mini band gap opening when $g=a/4$ and $g=3a/8$ and the corresponding reduction of the transmission range.
Furthermore, Fig.~\ref{fig4}(c) shows the variation with $g$ of the $N$-th and the $(N+1)$-th band intersection with the X point.
The phononic band gap is completely opened for $g=0$ and $g=a$, closes midway for $g=a/2$, and varies continuously and symmetrically between these points.
Since the gap is fully opened for $g=0$ and closed for $g=a/2$, and the glide parameter can be continuously tuned with periodicity $a$, a continously-tunable transmission filter is obtained.
Note that symmetry protection against backscattering of the guided waves is only achieved when the glide parameter $g=a/2$.

\bigskip

Summarizing the results above, the interface waves are protected by a class of topology relying on spatial symmetries and thus belong to crystalline topological insulators \cite{prodanBOOK2016}.
Whereas crystalline topological phases generally induce interface waves that have a gapped spectrum, because the interface breaks the corresponding spatial symmetries, the
glide symmetry of the interface ensures a gapless Dirac point at the X point of the Brillouin zone.

The importance of glide-reflection symmetry is further verified in the SM \bibnotemark[SM] by considering the oblique lattice instead of the square lattice.
It is specifically found that inversion symmetry \cite{liPRA2020} combined with the glide operation leads to a gapped spectrum, unlike GRS.
We note that glide-reflection symmetric waveguides have been considered before, e.g. for microwaves~\cite{quevedoIEEEJM2021} or acoustic waves~\cite{jankovicPRA2021}, but that waveguiding is in this case ensured by structural boundaries rather than by a phononic band gap.
The existence of a complete band gap without glide ($g=0$) is indeed essential to our result.
Defect photonic crystal waveguides possessing glide-reflection symmetry have also been discussed in terms of symmetry \cite{mockPRB2010} or low group velocity \cite{patilOE2022}, but without consideration of their topological properties.

The SM \bibnotemark[SM] further explores the resilience of the interface waves to disorder, a direct check of symmetry protection. It is observed numerically that they survive a position disorder of at least 5\% of the lattice constant and an inclusion diameter disorder of 10\%.

The discussion so far has been limited to the scalar case of acoustic waves and to a square lattice crystal of steel rods in water.
It is obvious, however, that the symmetry principles involved extend the existence of glide-reflection symmetric crystal waveguides to other material systems and lattices.
The SM specifically illustrates the cases of acoustic waves in a fluid with rigid inclusions and of vector elastic waves in a solid perforated with cylindrical holes or containing solid inclusions \bibnotemark[SM].
By virtue of the well-established analogies between acoustic/elastic waves and optical/electromagnetic waves~\cite{kushwahaPRL1993}, the transposition to photonic crystals is straightforward.
Other wave systems such as plasmonic crystals, gravity-capillary waves at the surface of water~\cite{laforgeNJP2019}, or solutions of the Schr\"odinger equation are likely to present similar properties too.
Indeed, one can start from any wide band gap artificial crystal and produce a glide-reflection symmetric interface within it. Then, when the glide parameter is exactly one half of the lattice constant, a gapless spectrum is obtained whereas the non-glided crystal has a completely gapped spectrum. Hence, any efficient artificial crystal that has been designed in the past, from seismic waves at the meter scale \cite{brulePRL2014} to phononic crystals for thermal transport control at the nanometer scale \cite{yangNL2014}, can be used as a starting basis to design a completely new topological glided structure supporting symmetry protected edge states.

On the practical side, the glide operation offers the opportunity to design a continuously varying transmission that can be changed from no transmission (for $g=0$) to full transmission through the phononic band gap (for $g=a/2$).
As a waveguide for transmission of information, the glide-reflection symmetric crystal waveguide offers wide bandwidth, single mode operation, and symmetry-protected backscattering immunity.

\section*{Acknowledgments}
The authors are grateful for support by the EIPHI Graduate School (Contract No. ANR-17-EURE-0002).


%

\end{document}